\documentclass[aps, prb,twocolumn,superscriptaddress, nofootinbib, showpacs]{revtex4-1}
\pdfoutput = 1

\usepackage{graphicx}
\usepackage{dcolumn}
\usepackage{bm}
\usepackage{dcolumn}
\usepackage{textgreek}
\usepackage{textcomp}

\newcommand{\textsubscript}[1]{$_{\text{#1}}$}

\def\BNFA{Ba\textsubscript{$1-x$}Na\textsubscript{$x$}Fe\textsubscript{$2$}As\textsubscript{$2$}}
\def\SNFA{Sr\textsubscript{$1-x$}Na\textsubscript{$x$}Fe\textsubscript{$2$}As\textsubscript{$2$}}
\def\BKFA{Ba\textsubscript{$1-x$}K\textsubscript{$x$}Fe\textsubscript{$2$}As\textsubscript{$2$}}
\def\BFA{BaFe\textsubscript{$2$}As\textsubscript{$2$}}
\def\NFA{NaFe\textsubscript{$2$}As\textsubscript{$2$}}
\def\AFA{\textit{A}Fe\textsubscript{$2$}As\textsubscript{$2$}}

\def\I4mmm{\textit{I}4/\textit{mmm}}

\def\hho{$\frac{1}{2} \frac{1}{2} 1$}
\def\hht{$\frac{1}{2} \frac{1}{2} 3$}

\begin{document}

\preprint{APS/123-QED}

\title{Detailed magnetic and structural analysis mapping a robust magnetic $C_4$ dome in  \SNFA}

\author{K.M. Taddei}
\email[corresponding author ]{ktaddei@anl.gov}
\affiliation{Department of Physics, Northern Illinois University, DeKalb, IL 60115, USA}
\affiliation{Materials Science Division, Argonne National Laboratory, Argonne, IL 60439, USA}
\author{J.M. Allred}
\affiliation{Materials Science Division, Argonne National Laboratory, Argonne, IL 60439, USA}
\affiliation{Department of Chemistry, University of Alabama, Tuscaloosa, AL  35487, USA}
\author{D.E. Bugaris}
\affiliation{Materials Science Division, Argonne National Laboratory, Argonne, IL 60439, USA}
\author{S.H. Lapidus}
\affiliation{Advanced Photon Source, Argonne National Laboratory, Argonne, IL 60439, USA}
\author{M.J. Krogstad}
\affiliation{Department of Physics, Northern Illinois University, DeKalb, IL 60115, USA}
\affiliation{Materials Science Division, Argonne National Laboratory, Argonne, IL 60439, USA}
\author{R. Stadel}
\affiliation{Department of Physics, Northern Illinois University, DeKalb, IL 60115, USA}
\affiliation{Materials Science Division, Argonne National Laboratory, Argonne, IL 60439, USA}
\author{H. Claus}
\affiliation{Materials Science Division, Argonne National Laboratory, Argonne, IL 60439, USA}
\author{D.Y. Chung}
\affiliation{Materials Science Division, Argonne National Laboratory, Argonne, IL 60439, USA}
\author{M.G. Kanatzidis}
\affiliation{Materials Science Division, Argonne National Laboratory, Argonne, IL 60439, USA}
\affiliation{Department of Chemistry, Northwestern University, Evanston, IL 60208, USA}
\author{S. Rosenkranz}
\affiliation{Materials Science Division, Argonne National Laboratory, Argonne, IL 60439, USA}
\author{R. Osborn}
\affiliation{Materials Science Division, Argonne National Laboratory, Argonne, IL 60439, USA}
\author{O. Chmaissem}
\affiliation{Department of Physics, Northern Illinois University, DeKalb, IL 60115, USA}
\affiliation{Materials Science Division, Argonne National Laboratory, Argonne, IL 60439, USA}

\date{\today}

\begin{abstract}
The recently discovered $C_4$ tetragonal magnetic phase in hole-doped members of the iron-based superconductors provides new insights into the origin of unconventional superconductivity. Previously observed in Ba\textsubscript{1-x}\textit{A}\textsubscript{x}Fe\textsubscript{2}As\textsubscript{2} (with $A =$ K, Na), the $C_4$ magnetic phase exists within the well studied $C_2$ spin-density wave (SDW) dome, arising just before the complete suppression of antiferromagnetic (AFM) order but after the onset of superconductivity. Here, we present detailed x-ray and neutron diffraction studies of \SNFA\ ($0.10 \leq\ x \leq\ 0.60$) to determine their structural evolution and the extent of the $C_4$ phase. Spanning $\Delta x\sim 0.14$ in composition, the $C_4$ phase is found to extend over a larger range of compositions, and to exhibit a significantly higher transition temperature, $T_r \sim 65$K, than in either of the other systems in which it has been observed. The onset of this phase is seen near a composition ($x \sim 0.30$) where the bonding angles of the Fe\textsubscript{$2$}As\textsubscript{$2$} layers approach the perfect $109.46^\circ$ tetrahedral angle. We discuss the possible role of this return to a higher symmetry environment for the magnetic iron site in triggering the magnetic reorientation and the coupled re-entrance to the tetragonal structure. Finally, we present a new phase diagram, complete with the $C_4$ phase, and use its observation in a third hole-doped 122 system to suggest the universality of this phase.    
\end{abstract}


\maketitle

\section{\label{sec:intro}Introduction}
The recent discovery in \BNFA\ of a new antiferromagnetic phase, which restores tetragonal symmetry at temperatures below the transition to the more commonly observed orthorhombic antiferromagnetism, has important implications for the nature of unconventional superconductivity in the iron pnictides.\cite{Avci2014} In these systems, the structural and magnetic phase transitions are strongly coupled, and two major schools of thought have emerged favoring either magnetic fluctuations (Ref~\onlinecite{Eremin2010, Castellan2011, Fernandes2013}) or orbital ordering (Ref~\onlinecite{Lv2010, Yi2011, Chu2012, Kuo2013}) as the primary driving force.  Discriminating between these two models is complicated by the fact that magnetoelastic coupling ensures that the onset of one order parameter triggers the other,\cite{Rotter2008BN} and indeed the structural and magnetic phase transitions are coincident and first-order in many of the iron pnictides.\cite{Avci2011, Avci2012, Avci2013, Allred2014}  However, a resolution of this issue will provide strong constraints on the origin and symmetry of the superconducting order parameter.\cite{Fernandes2014}

Magnetic order in \BFA\ and related \lq 122\rq\ structures consists of antiferromagnetic stripes, in which iron spins within each plane are ferromagnetically aligned along one iron-iron bond direction and antiferromagnetically aligned along the orthogonal bond. The magnetic moments are aligned within the plane parallel to the antiferromagnetic bonds. This magnetic structure breaks the four-fold symmetry of the iron atom square lattice and is accompanied by a reduction in the symmetry of the atomic lattice from tetragonal to orthorhombic, \textit{i.e.}, from the \I4mmm\ to \textit{Fmmm} space groups. In the following, we refer to this as the $C_2$ phase. The transition to the new magnetic phase, which restores four-fold \I4mmm\ symmetry to the atomic lattice and so is referred to as the $C_4$ phase, occurs at temperatures ($T_r$) below the $C_2$ transition. The magnetic Bragg peaks have the same reciprocal space indices in both the $C_2$ and $C_4$ phases,\cite{Avci2014} although the spins in the $C_4$ phase are oriented parallel to the $c$ axis.\cite{Wasser2015} One way to achieve this is for the magnetic order to consist of a double-\textbf{Q} structure, comprising the superposition of stripes along both the $x$ and $y$ directions within the same domain. It is, also, possible to construct models of orbital order that are consistent with a tetragonal space group, but they are incompatible with double-\textbf{Q} magnetic order.\cite{Khalyavin2014}

Recently, M\"ossbauer data combined with high-resolution neutron and x-ray diffraction on a new compound, Sr\textsubscript{$0.63$}Na\textsubscript{$0.37$}Fe\textsubscript{$2$}As\textsubscript{$2$}, has conclusively demonstrated that the $C_4$ magnetic structure is a double-\textbf{Q} spin-density wave (SDW).\cite{Allred2015N} This sample, which is a member of the series that forms the subject of this paper, exhibits a transition from the paramagnetic tetragonal phase to the $C_2$ phase at $\sim$ 105~K and then a strongly first-order transition to the $C_4$ phase at about 65~K.  The M\"ossbauer data unequivocally demonstrates that, in the $C_4$ phase, 50\% of the iron sites are non-magnetic and 50\% have double the moment measured in the $C_2$ phase. This is confirmation that the $C_4$ phase arises from the constructive and destructive interference of two orthogonal SDWs. More details can be found elsewhere.\cite{Allred2015N,Avci2014}

The observed double-\textbf{Q} structure requires the transfer of magnetization density from non-magnetic to magnetic sites. This is inconsistent with localized models of magnetic moments with fixed amplitudes on each site. It is, however, consistent with more weakly coupled models, in which a modulation of the itinerant electron spin density is caused by quasi-nesting features of the Fermi surface. In this scenario, interband interactions between the hole pockets at the zone center and the electron pockets at the zone boundary generate strong magnetic fluctuations along both the $x$ and $y$ axes. In the $C_2$ phase, there is a breaking of Ising symmetry by nesting along one or other directions, whereas in the $C_4$ phase, there is a simultaneous nesting along both directions, restoring four-fold symmetry. The $C_4$ phase is predicted to be stabilized by an increasing mismatch in the size of the hole and electron pockets.\cite{Avci2014} The spin reorientation follows from symmetry considerations in the presence of strong spin-orbit coupling.\cite{Allred2015N, Wasser2015, Mallett2015muon} 

The $C_4$ transition is higher in \SNFA\ than in the other compounds, in which it has been observed. This indicates that the tetragonal phase is more stable in this series, so we mapped out the entire phase diagram in this work. We report the synthesis of high quality \SNFA\ samples with $x$ up to a nominal composition of 0.6.  Samples beyond the $x = 0.6$ composition are not investigated here because they are beyond the region of $C_4$ stability and are expected to show purely superconducting transitions with reduced $T_c$'s down to $\sim11$ K for the metastable \NFA\ compound.\cite{Cortes2011,Shinohara2015,Todorov2010} The compositional range chosen for this study allows us to fully focus on the region of phase coexistence and phase competition among diverse ground states.  Universality of the $C_4$ phase in the hole-doped pnictides is fully established by this study with \SNFA\ being the third known series to show the existence of this novel magnetic phase after \BNFA\ and \BKFA .\cite{Avci2014, Allred2015, Bohmer2015}  We will delineate the relatively large region of the $C_4$ phase with $T_r$'s peaking at $\sim 65$ K.  The relative stability of the samples in air coupled with the $C_2$, $C_4$, and SC phase competition will undoubtedly provide strong clues for solving the unconventional nature of superconductivity in the pnictide superconductors.

The organization of this report will be briefly described here to reduce ambiguity as to when, in the following discussions, the different phases of this system are being considered. Section~\ref{sec:exp} will detail the synthesis and preliminary characterization of the samples which were the basis of this study. Section~\ref{sec:resdis} will present and discuss the results of our neutron and x-ray diffraction experiments. All results and final conclusions will be summarized in Section~\ref{sec:conc} where a complete phase diagram is presented. The introduction of the new $C_4$ magnetic phase is not pertinent to all results discussed and at times needlessly complicates descriptions.  Therefore, discussion of this phase will be reserved until Subsection~\ref{subsect:C4} and the following sections, with brief allusions to its existence and effects on the structure being made only where necessary in preceding sections.

\section{\label{sec:exp}Experimental Details}
\subsection{\label{subsect:synth}Synthesis of \SNFA}
Twenty-three compositions were synthesized, as polycrystalline powders, with nominal sodium contents $x =$ 0.10, 0.20, 0.25, 0.26, 0.28, 0.29, 0.30, 0.32, 0.34, 0.35, 0.36, 0.37, 0.38, 0.39, 0.40, 0.42, 0.44, 0.45, 0.5, and 0.6.  Duplicate samples have been prepared at different times for the diverse diffraction experiments.  Samples prepared for neutron diffraction were approximately 5g each while samples used for synchrotron x-ray diffraction were only $\sim0.5$g each.  Despite the overall agreement in structural and physical properties, subtle differences are occasionally observed due to the disparate sample size and the complex synthesis procedure.  Handling of all the starting materials was performed in an M-Braun glovebox under an inert Ar atmosphere with less than 0.1 ppm of H\textsubscript{2}O and O\textsubscript{2}.  Starting Sr (Aldrich, 99.9\%) and Fe (Alfa Aesar, 99.99+\%) elements were used as received.  Small pieces of Na free of oxide coating were trimmed from large lumps (Aldrich, 99\%).  Granules of As (Alfa Aesar, 99.99999+\%) were ground to a coarse powder prior to use.  Precursor binary materials SrAs, NaAs, and Fe\textsubscript{$2$}As were synthesized from stoichiometric reactions of the elements at 800\textdegree C, 350\textdegree C, and 700\textdegree C, respectively.  Polycrystalline samples of \SNFA\ were prepared from stoichiometric mixtures of SrAs, NaAs, and Fe\textsubscript{$2$}As, which were ground thoroughly with a mortar and pestle, and loaded in alumina crucibles.  The alumina crucibles were sealed in Nb tubes under Ar, which were further sealed in quartz tubes under vacuum.  The reaction mixtures were subjected to multiple heating cycles between 850-1000\textdegree C for durations less than 48 h (to minimize loss of Na by volatilization).  The samples underwent grinding by mortar and pestle between heating cycles in order to homogenize the composition.  Following the final heating cycles, the samples were quenched in air from the maximum temperature rather than being allowed to cool slowly.  Initial characterization of the dark gray powders was conducted by laboratory magnetization measurements at 0.1 Oe on a home-built SQUID magnetometer to determine $T_c$.  

\begin{table*}
	\caption{\label{table:1}Fitted composition and structural and magnetic transition temperatures.  Fit compositions were determined through the use of a Vegard's law-like behavior of the \textit{a} lattice parameter. Samples which show significant departures from the nominal composition had been post-annealed with tiny excess amounts of Sr or Na in order to improve the quality of the superconducting transition.  $T_c$'s were determined as the onset of the diamagnetic response through the intersection of the linear fits of the curve before and during the transition.  Due to the suppression of the orthorhombic and accompanying magnetic transitions ($ T_s$ and $T_N$) by the re-entrant $C_4$ phase, these transitions were determined using the same technique. In samples which showed purely the $C_2$ magnetic phase, a power law fit ( $M(\delta) \propto (T_{N(s)}-T)^{\beta_{N(s)}})$ was used to determine the transition temperatures and critical exponents.  $T_{r,s}$ and $T_{r,N}$ denote the structural re-entrance and magentic reorientation transitions respectively.}
	\begin{ruledtabular}
	\begin{tabular}{lccdrdrccc}
			& $x_{nom}$ & $x_{fit}$ & \multicolumn{1}{c}{\textrm{\textit{T\textsubscript{c}}}} & \multicolumn{1}{c}{\textrm{\textit{T\textsubscript{N}}}} & \multicolumn{1}{c}{\textrm{\textbeta\textit{\textsubscript{N}}}} & \multicolumn{1}{c}{\textrm{\textit{T\textsubscript{s}}}} & $\beta_s$ & $T_{r,s}$ & $T_{r,N}$\\
			 \hline
	x-ray 	& 	   &      &      &  		&			&        &			&       &		 \\
		 	& 0.10 & 0.12 &      &  		&   		& 182(3) & 			&       &		 \\
		 	& 0.20 & 0.19 & 7    &  		&   		& 162(3) & 			&       &		 \\
		 	& 0.25 & 0.27 & 7    &  		&   		&        & 			&       &		 \\
		 	& 0.26 & 0.32 & 10   &  		&   		& 115(3) & 			& 50(3) &		 \\
		 	& 0.28 & 0.30 & 11   &  		&   		& 128(3) & 			& 20(4) &		 \\
		 	& 0.30 &      & 8    &  		&   		&        & 			&       &		 \\
		 	& 0.32 & 0.29 & 11   &  		&   		&        & 			& 15(4) &		 \\
		 	& 0.35 & 0.35 & 12   &  		&   		& 105(3) & 			&       &		 \\
		 	& 0.36 & 0.28 & 8    &  		&   		&        & 			&       &		 \\
		 	& 0.37 & 0.34 & 10   &  		&   		& 112(4) & 			& 65(3) &		 \\
		 	& 0.40 & 0.40 & 16   &  		&   		& 103(4) & 			& 65(3) &		 \\
		 	& 0.42 &      & 23   &			&   		& 70(3)  & 			&       &		 \\
		 	& 0.50 & 0.51 & 36   &  		&   		&        & 			&       &		 \\					
		 	& 0.60 & 0.59 & 34	 &  		&   		&        & 			&       &		 \\				
	neutron &      &      &      &    		& 			&        & 			&       &		 \\
			& 0.29 & 0.29 & 9	 & 139(1)	& 0.32(1)	& 139(1) & 	0.24(1)	& 		&		 \\
			& 0.32 & 0.37 & 11	 & 		    &   		&   	 &			&		&		 \\
			& 0.35 & 0.38 &	12	 & 108(1)	& 0.53(7)	&   	 &			&		& 66(1)  \\
			& 0.37 & 0.36 & 11	 & 115(1)	& 0.42(7)	& 112(2) &			& 67(3)	& 65(1)  \\
			& 0.37 & 0.41 & 21	 &   		&   		&   	 &			&		&		 \\
			& 0.40 & 0.43 &	24	 & 75(1)	& 0.48(5)	&   	 &			&		&		 \\
			& 0.42 & 0.42 &	22	 & 77(2)	& 0.17(6)   &   	 &			&		&		 \\
			& 0.45 & 0.45 & 37	 & 56(2)	& 0.20(10)	&        &			&		&		 \\
			& 0.48 & 0.48 &	32   &  		&   		&   	 &			&		&		 \\						
	\end{tabular}
	\end{ruledtabular}
\end{table*}

\begin{figure}
	\includegraphics[width=\columnwidth]{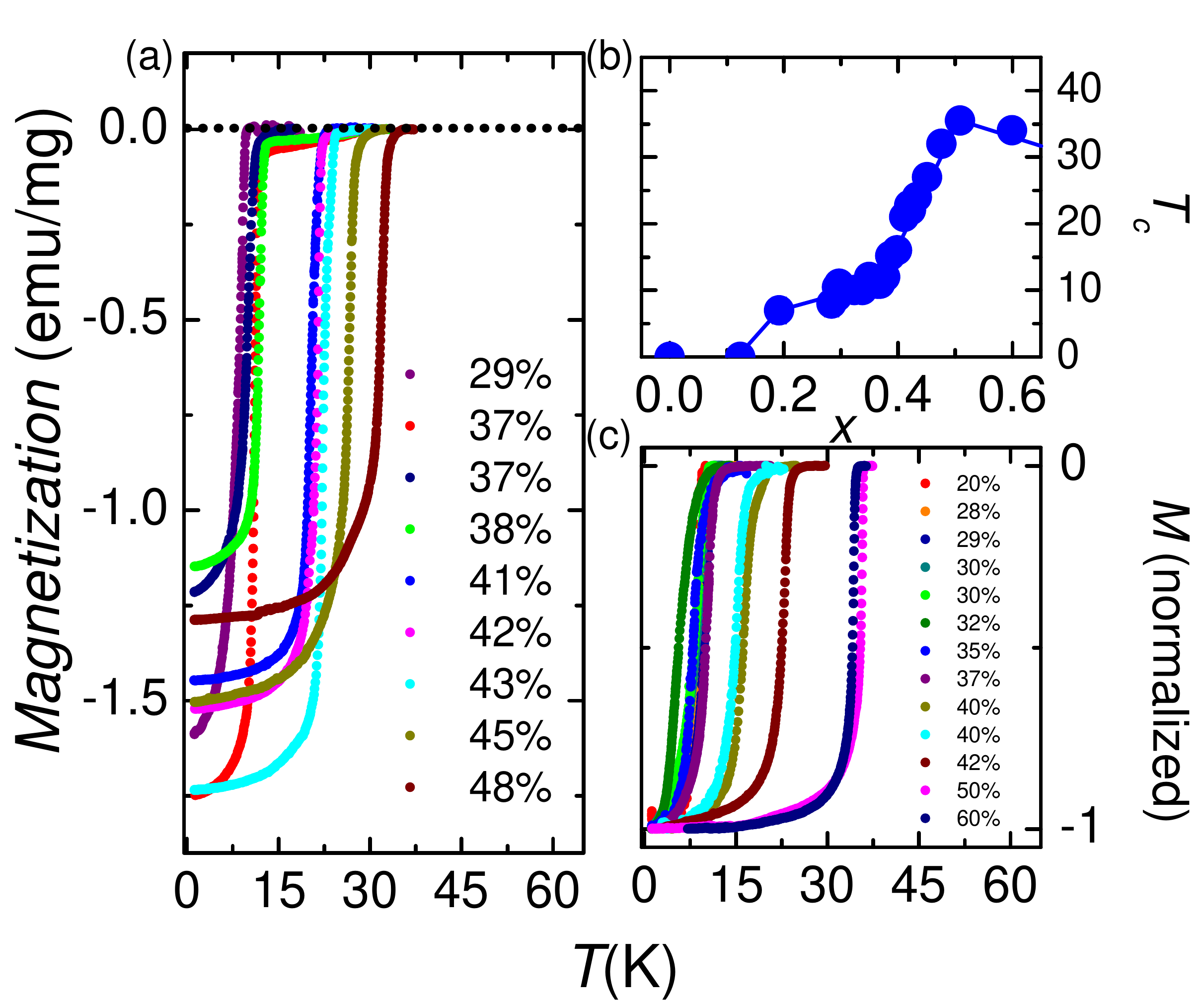}	
	\caption{\label{fig:one} (Color online) The superconducting transitions for all measured powders. (a) Magnetization normalized to mass of the large samples used for neutron diffraction, all samples show bulk superconductivity. The variation in diamagnetic response is consistent with small FeAs impurities seen in diffraction patterns. (b) Superconducting transition as a function of composition defining the superconducting dome. (c) Normalized magnetization curves for small batches prepared for x-ray measurements (Due to the high number of similar compositions, some curves are not visible because of overlap).}	
\end{figure}

\subsection{\label{subsec:char}Sample Characterization}

Time-of-flight (TOF) neutron powder diffraction experiments were performed using POWGEN at the Spallation Neutron Source (SNS) of Oak Ridge National Laboratory (ORNL).  Magnetic order parameter measurements using the same powder samples were performed at the triple-axis beamline HB-1A of the High Flux Isotope Reactor (HFIR) where special attention was given to collecting data for the inherently weak magnetic reflections. High resolution synchrotron x-ray data were collected at beamline 11BM-B of the Advanced Photon Source (APS) at Argonne National Laboratory (ANL).  Detailed structural analyses were performed using the Rietveld method as implemented in the GSAS and EXPGUI software suite.\cite{Toby2001,Larson2004} In the final refinement cycles, all parameters were allowed to vary, including fractional coordinates, thermal factors, site occupancies, background, absorption correction, and peak shape profiles. Back-to-back exponentials convoluted with a pseudo-Voigt and employing microstrain broadening were used to model the TOF peak shape profile.\cite{Stephens1999}  A pseudo-Voight peak shape profile function 3 was used with the synchrotron data. 

In order to ensure the highest quality samples, the annealing was continually monitored by magnetization measurements after each heat treatment.  Repeated grinding and annealing steps during the synthesis were found necessary not only to ensure the chemical homogeneity of the sample but also to produce sharp single superconducting transitions as shown in Figures~\ref{fig:one}\textit{a} and \textit{c}.  The final samples are found to remain stable when exposed to air for periods of several days in clear contrast with the metastable nature of their hole-doped analogs such as the air sensitive \BNFA\ series.

\section{\label{sec:resdis} Results and Discussion}

\subsection{\label{subsec:sc}Superconductivity}

As shown in Figure~\ref{fig:one}\textit{b}, superconducting samples covering a significant portion of the expected superconducting dome were produced with $T_c$ peaking at $\sim$ 36 K for the $x = 0.5$ composition.  We also note the somewhat shallow left tail of the dome extending well between $x \sim 0.2$ and $x \sim 0.4$ making it obvious that the analysis of compositions within this range cannot rely solely on the measured $T_c$ but must also  include the refined structure and lattice properties. Table \ref{table:1} shows the nominal composition compared to the composition determined from Vegard's law-like fits performed using the linear composition dependence of the \textit{a} lattice parameter at room temperature.  All references to the sample composition will invoke the corrected $x_{fit}$ composition.

\subsection{\label{subsect:dopdep}Structural Properties and Comparison to Other Hole-Doped 122 Materials}
\begin{figure}
\includegraphics[width=\columnwidth]{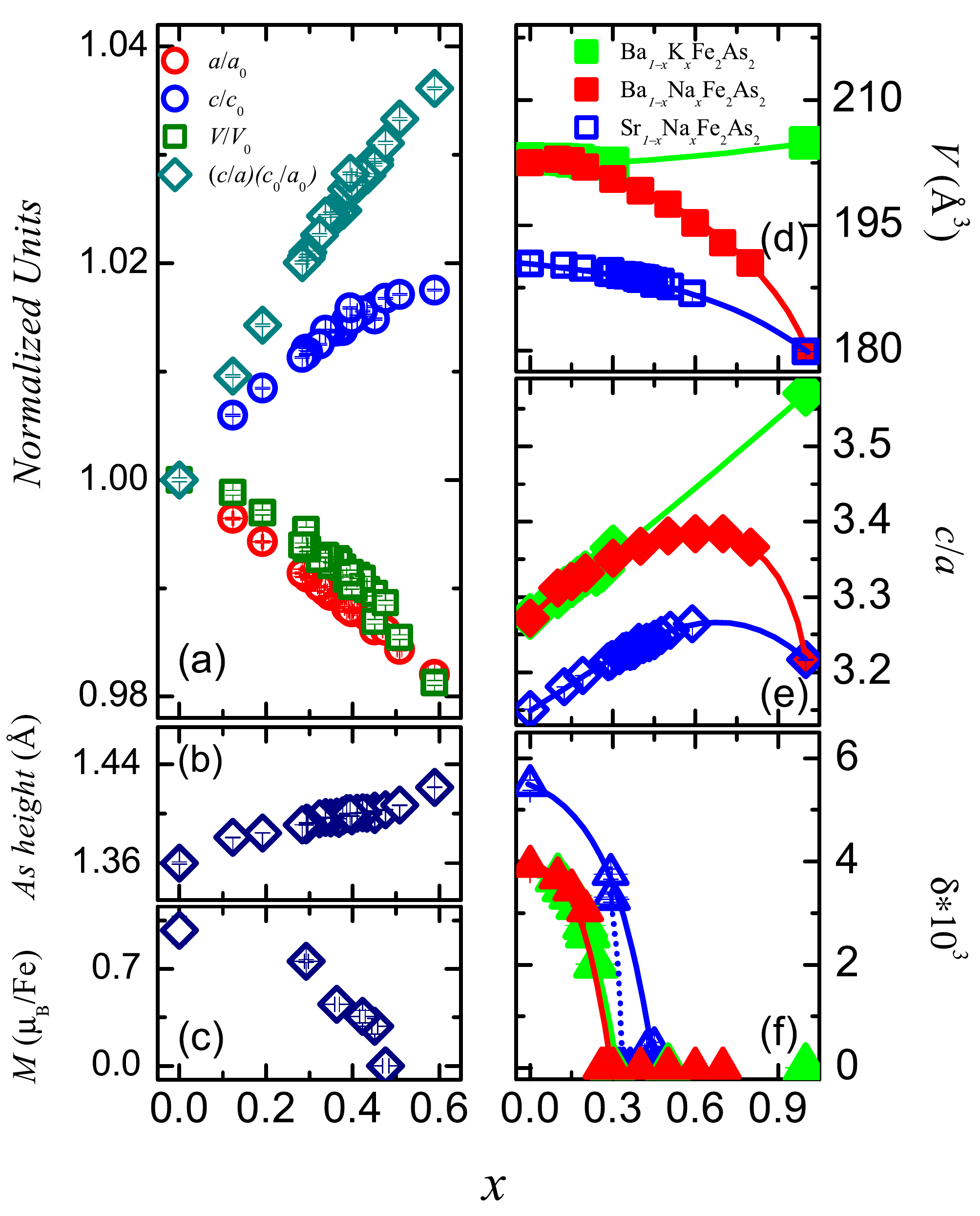}
\caption{\label{fig:2} (Color online) Doping dependence of: (a) the lattice parameters ($a, c, V,$ and $c/a$) normalized to the parent SrFe\textsubscript{$2$}As\textsubscript{$2$} structure, (b) the distance of the arsenic atom site from the Fe-Fe square lattice, (c) and the magnetic moment. The doping dependence of the \SNFA\ system's (d) volume ($V$), (e) lattice anisotropy ($c/a$) and (f) orthorhombic order parameter ($\delta$) as compared to the other two members of the hole doped 122 system \BNFA\ and \BKFA .  Parent compound's lattice parameters' and arsenic position are taken from reference \onlinecite{Alireza2009} while its magnetic moment is taken from reference \onlinecite{Mohanta2013}. Panels d-f are made with data originally published in references \onlinecite{Avci2013} and \onlinecite{Avci2011}. Lines have been added to d-f as guides to the eye. While the solid blue line in (f) was estimates the behavior of $\delta$ in the absence of the $C_4$ phase the dotted line shows the actual behavior.}
\end{figure}

The substitution of Na on the Sr site causes two main changes from the parent compound which must be considered in understanding the doping dependence of the structure: first, the Na\textsuperscript{+} ion contributes one less electron than Sr\textsuperscript{2+} and so decreases the oxidation state of Fe (this shift in the charge of the Fe\textsubscript{$2$}As\textsubscript{$2$} tetrahedron greatly affects the geometry of these layers), and second, the smaller ionic radius of Na requires the lattice of the material to progressively accommodate the size mismatch as more Sr is replaced by Na. Figure~\ref{fig:2}\textit{a} shows the room temperature lattice parameters as a function of doping normalized to the parent compound.  Both the volume (\textit{V}) and either direction along the tetragonal basal plane are seen to decrease nearly linearly with increasing Na doping, demonstrating the combined effect of these two mechanisms.  The change in the \textit{a} axis between the parent compound and our highest doped sample of $x = 0.59$ is approximately -1.8\%.  Surprisingly, the \textit{c} axis is seen to expand by a compensating +1.75\%.  In order to understand this feature the behavior of the FeAs layer must be considered. Fig~\ref{fig:2}\textit{b} shows the As site's distance from the Fe plane. As the basal plane contracts due, partially, to the smaller size of the Na\textsuperscript{+} ion and more significantly to the increased oxidation state of Fe the well-known relative rigidity of the Fe-As bond length causes the As to be pushed higher above and below the plane, consequently, leading to the observed expansion of the unit cell along the \textit{c} direction (see Section~\ref{subsect:intParam} for a more detailed analysis of the internal parameters). \cite{Avci2013}  

The \textit{c/a} ratio can be used as a measure of the lattice anisotropy, and it is seen to monotonically increase with doping (Figure~\ref{fig:2}\textit{e}).  Interestingly, as the anisotropy and interlayer distance increase (indicated by the increasing \textit{c} axis), the magnitude of the magnetic ordering decreases (Fig~\ref{fig:2}c).  A similar behavior is observed in all members of the hole-doped compounds and may be naively attributed to weakening interlayer magnetic correlations as the neighboring layers become increasingly distant. \cite{Avci2011,Avci2013} Comparing the \textit{c/a} ratio to the volume it can be seen that the contraction along the basal plane has a larger effect on the unit cell volume than does the expansion along \textit{c} and so the volume shrinks in accordance with expectations as the lattice changes to accommodate the smaller Na atom.

Comparisons with the other prominent hole-doped 122 systems are presented in Fig~\ref{fig:2}\textit{d-e}. While the two Ba compounds (\BKFA\ and \BNFA ) have similar structural properties for compositions $x \leq\ 0.4$ due to their sharing a common parent compound, the much smaller Sr atom causes the volume of this series to be significantly less than for the Ba analogue (Fig~\ref{fig:2}d). However, replacing Ba with a smaller \textit{A} site ion does not similarly reduce both the \textit{a} and \textit{c} lattice parameters as can be seen in Figure~\ref{fig:2}e where the anisotropy ratio is 4\% less for the Sr system - while the change from Ba to Sr causes only a tiny 0.8\% change in the \textit{a} axis (from 3.95537 \AA\ to 3.9243 \AA), the \textit{c} axis changes by a relatively significant 5\% (from 12.9424 \AA\  to 12.3644 \AA).  This reduction in $c/a$ has profound effects on the internal parameters as will be discussed in section~\ref{subsect:intParam}. 

It is worth noting that the non-linear doping dependence observed in the volume of \BNFA\ is also present in the Sr system (Figure~\ref{fig:2}\textit{e}). We previously ascribed this behavior to the stresses placed on the lattice by substitution of the significantly smaller Na\textsuperscript{+} ion.\cite{Avci2013} While the change in the oxidation state of Fe is the dominant affect in the underdoped region (as evidenced by the nearly identical features of \BNFA\ and \BKFA\ for $ x < 0.4$), the internal stresses caused by the smaller size of the Na atom become more significant with higher doping and eventually lead to the formation of the metastable \NFA .   This shared end-member has a significantly reduced \textit{c} axis but a similar \textit{a} axis leading to a smaller volume of $\sim180\text{\AA}^3$. We propose here the same mechanism to describe the similar behavior of the Sr system albeit somewhat mitigated by the smaller size mismatch between the SrFe\textsubscript{$2$}As\textsubscript{$2$} and \NFA\ end-members. \cite{Todorov2010} 

Figure~\ref{fig:2}\textit{f} shows the orthorhombic order parameter for the three systems at 10K.  The orthorhombic splitting is both larger and persists to higher dopant concentrations in \SNFA\ than for either of the other two systems. Recently, similar behavior was observed in the related intercalated iron selenide \lq 122\rq\ family of superconductors (\textit{A}\textsubscript{$x$}Fe\textsubscript{$1-y$}Se\textsubscript{$2$} with \textit{A} = Na, K, Rb, or Cs) where it was suggested that the strength and ordering temperature of the magnetic phase was dependent on the size of the intercalating ion and consequently the spacing between the tetrahedral Fe\textsubscript{$2$}Se\textsubscript{$2$} layers. \cite{Taddei2015} For the hole-doped iron arsenide systems being considered here a similar dependence is seen where $T_N$ decreases with increasing ionic radius ($r_A$) as monitored by the \textit{a} lattice parameter, with $T_N = 205, 140$K and $a = 3.9243(1), 3.9625(1)$ \AA\  for $A$ = Sr and Ba respectively. \cite{Tegel2008,Wu2008,Krellner2008}  While the magnetic transition in the iron selenides is not strongly coupled to a structural transition, the strong magneto-elastic coupling in the hole-doped 122 iron arsenides, where magnetism is the primary order parameter, suggests it is likely that the smaller lattice of the Sr system allows for larger magnetic interactions between neighboring iron sites and so enhances the behavior of the structural and magnetic phase transitions.  This behavior would then account for the higher ordering temperature and larger ferromagnetic and antiferromagnetic interactions along the \textit{b} and \textit{a} lattice parameters, respectively (see section~\ref{subsect:Tmag}),  and causing, through the strong magneto-elastic coupling, a correspondingly larger structural distortion \cite{Avci2011}. 

\begin{figure}
\includegraphics[width=\columnwidth]{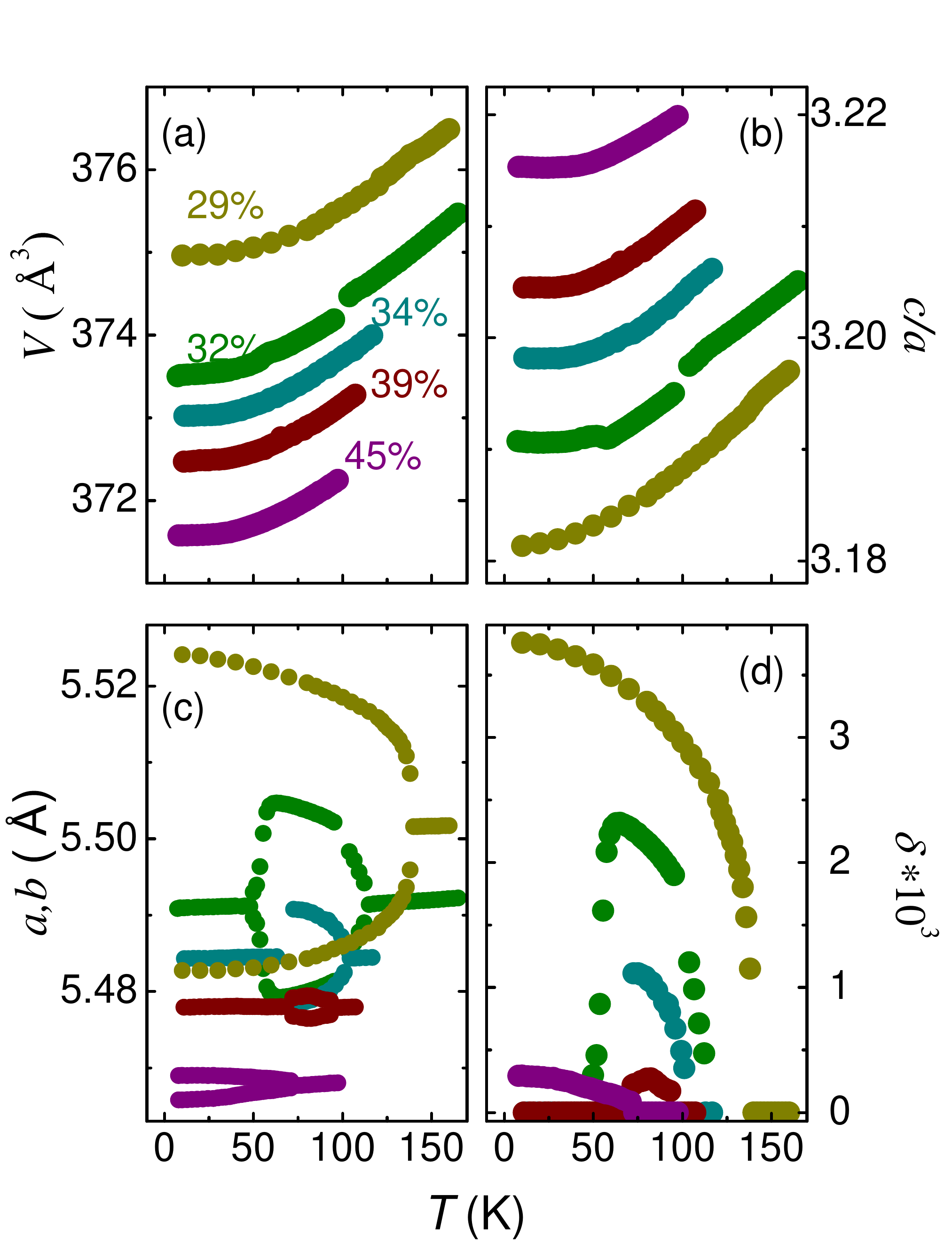}
\caption{\label{fig:3} (Color online) Temperature dependence for the: (a) volume ($V$), (b) lattice anisotropy, (c)  \textit{a} and \textit{b} lattice parameters and (d) orthorhombic order parameter ($\delta$) for compositions $x =$ 0.29, 0.32, 0.34, 0.39 and 0.45 determined from Rietveld refinements using synchrotron x-ray and spallation source neutron data.}
\end{figure}

\subsection{\label{subsec:Tstruct}Temperature Dependence of Structural Parameters}

\begin{figure}[b]
\includegraphics[width=\columnwidth]{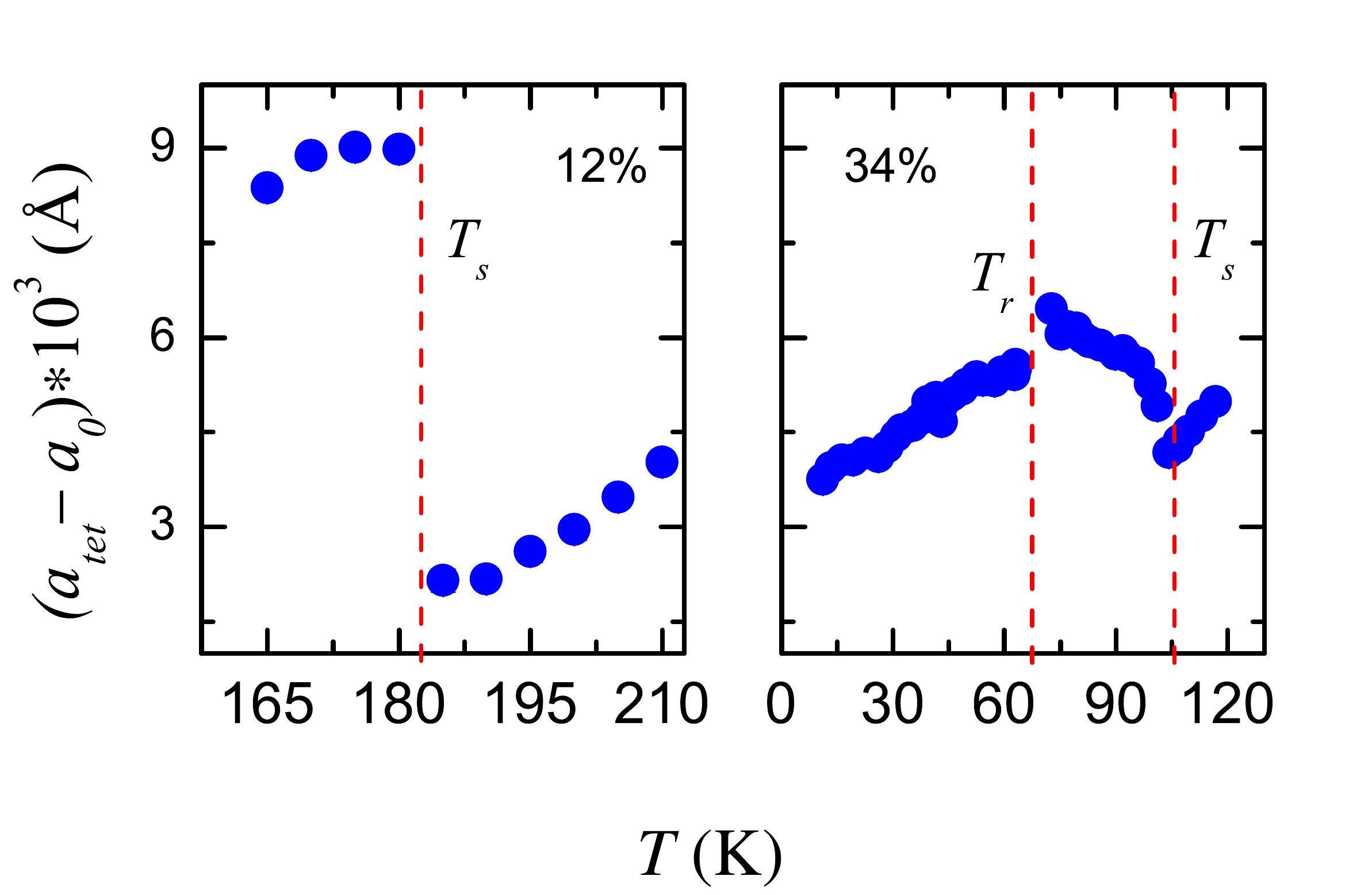}
\caption{\label{fig:4} (Color online) First-order-like nature of discontinuity in \textit{a} lattice parameter at $T_s$($T_N$) and $T_r$ (marked approximately by dotted red lines) for representative 13\% and 34\% samples.  $a_{tet}$ calculated in orthorhombic structure as $a_{tet} = \sqrt{a_{oth}^2+b_{orth}^2}/2 $.  Plotted value has been scaled by three orders of magnitude after subtraction from $a_0 = 3.9089$ and $3.8776 $ \AA\ for the 12\% and 34\% samples respectively.}
\end{figure}

\begingroup
\squeezetable
\begin{turnpage}
\begin{table*}
	\caption{\label{table:2}Results of Rietveld refinements from room temperature and 10K x-ray (11BM-B) and neutron (POWGEN)data. Listed composition is fit composition (see Table~\ref{table:1}). 300K data structure has \I4mmm\ symmetry.  10K model has either $C_2$ \textit{Fmmm} or $C_4$ \I4mmm\ symmetry depending on composition.  10K data modeled with $C_4$ symmetry has \textit{a} and \textit{b} lattice parameters scaled by $\sqrt{2}$.}
	\begin{ruledtabular}
	\begin{tabular}{lclclllllcllc}
			&  $x_{fit}$  &  \multicolumn{1}{c}{\textrm{\textit{a} (\AA)}}   &  \textit{b}(\AA)  & \multicolumn{1}{c}{\textrm{\textit{c}(\AA)}}     & \multicolumn{1}{c}{\textrm{\textit{V} (\AA\textsuperscript{3})}} & \multicolumn{1}{c}{$z_{As}$}  & \multicolumn{1}{c}{\textrm{Fe-As (\AA)}} & \multicolumn{1}{c}{\textrm{Sr-As (\AA)}} & Sr-As (\AA) &  \multicolumn{1}{c}{$\alpha_{1}$  (\textdegree)}   &  \multicolumn{1}{c}{$\alpha_{2}^{''}$ (\textdegree)} &  $\alpha_{2}^{''}$  (\textdegree) \\
			 \hline
	300K 	        &           &		 &	      &		    &				   &		&             &	 	    &		  &			    &				    &	--			     \\
		 	&  0.12     & 3.91031(6) &  --        & 12.4382(3)  & 190.188(7)                   & 0.3610(6)  & 2.3936(3)   &	 3.2610(3)  &  --	  &  109.54(2)		    &  109.44(1)		    &	--			     \\
		 	&  0.29     & 3.8920(2)  &  --        & 12.513(1)   & 189.541(2)		   & 0.3611(6)  & 2.3920(5)   &	 3.2546(4)  &  --	  &  108.89(3)		    &  109.76(2)		    &	--			     \\
		 	&  0.32     & 3.88561(6) &  --        & 12.5193(2)  & 189.016(4)                   & 0.3614(5)  & 2.3922(2)   &	 3.2490(2)  &  --	  &  108.61(2)		    &  109.904(8)		    &	--			     \\
		 	&  0.34     & 3.8839(4)	 &  --        & 12.535(3)   & 189.082(5)		   & 0.3614(8)	& 2.3921(1)   &	 3.2495(1)  &  --	  &  108.54(1)		    &  109.94(5)		    &	--			     \\
		 	&  0.37     & 3.8834(2)  &  --        & 12.5342(9)  & 189.024(2)		   & 0.3615(7)	& 2.3921(5)   &	 3.2489(5)  &  --	  &  108.53(4)		    &  109.95(2)		    &	--			     \\
		 	&  0.45     & 3.8728(1)  &  --        & 12.5623(9)  & 188.412(2)		   & 0.3616(6)	& 2.3880(3)   &	 3.2415(3)  &  --	  &  108.25(2)		    &  110.09(1)		    &	--			     \\
		 	&  0.48	    & 3.8697(7)  &  --        & 12.572(3)   & 188.253(8)		   & 0.3616(8)	& 2.390(1)    &	 3.243(1)   &  --	  &  108.11(9)		    &  110.16(4)		    &	--			     \\
		 	&  0.59     & 3.85373(5) &  --        & 12.5810(3)  & 186.843(1)		   & 0.363(5)	& 2.3944(4)   &	 3.2245(4)  &  --	  &  107.17(3)		    &  110.63(1)		    &	--			     \\
	10K             &           &		 &	      &		    &				   &		&  	      &	            &		  &			    &  				    &				     \\
			&  0.29     & 5.5189(2)	 & 5.4776(2)  &	12.369(5)   & 373.91(3)			   & 0.3613(6)	&  2.3819(4)  &  3.2318(4)  & 3.2494(4)	  &  109.40(3)    	    &  109.20(4) 		    &  109.81(1)		     \\
			&  0.32     & 5.48669(8) & --	      &	12.3788(3)  & 372.649(1)		   & 0.3615(4)	&  2.3775(2)  &  3.2380(2)  & --	  &  109.36(1)	            &  109.528(7)		    &  --			     \\
			&  0.34     & 5.4843(1)	 & --	      &	12.4020(4)  & 373.014(2)		   & 0.3618(5)	&  2.3824(4)  &  3.2349(3)  & --	  &  108.96(3)	            &  109.73(1) 		    &  --			     \\
			&  0.37     & 5.4857(3)	 & --	      &	12.398(1)   & 373.101(6)		   & 0.3619(8)	&  2.3820(7)  &  3.2358(6)  & --	  &  109.03(4)	            &  109.69(2) 		    &  --			     \\
			&  0.45     & 5.4703(9)	 & 5.4669(8)  &	12.425(2)   & 371.57(1)			   & 0.3619(8)	&  2.3808(3)  &  3.2275(3)  & 3.2293(3)	  &  108.56(2)	            &  109.90(1) 		    &  109.96(1)		     \\
			&  0.48     & 5.4626(7)	 & --	      &	12.433(2)   & 370.99(1)			   & 0.362(1)	&  2.3814(8)  &  3.2251(7)  & --	  &  108.39(5)	            &  110.01(3) 		    &  --	 		     \\
						
	\end{tabular}
	\end{ruledtabular}
\end{table*}
\end{turnpage}
\endgroup

At 205K, SrFe\textsubscript{2}As\textsubscript{2} undergoes the same \I4mmm\ to \textit{Fmmm} symmetry breaking as the Ba-122 system. \cite{Kumar2008,Krellner2008,Yan2008,Shinohara2015}  This transition breaks the structure's tetragonal symmetry through a structural distortion which causes the reorientation of the unit cell to a $\sqrt{2} \times \sqrt{2} \times 1$ supercell with the \textit{a} and \textit{b} axes no longer being symmetry equivalent.\cite{Rotter2008BN}

Figure~\ref{fig:3} shows the lattice's temperature dependence for a representative selection of compositions.  The \textit{c/a} and the mostly featureless volume plots (Fig~\ref{fig:3}\textit{a-b}) with only barely observable volume anomalies at $T_s$ indicate that the phase transition in this system is only weakly first-order, as will be demonstrated later in this section.\cite{Barzykin2009,Cano2010} The unit cell volume for all compositions shows the expected nearly linear dependence on temperature until $\sim$40K at which point the volume of the unit cell becomes effectively constant as is typical in these materials.  This behavior can also be seen in the \textit{c/a} plot as the lattice anisotropy decreases with falling temperature before reaching a minimum value at $\sim$40K. As discussed in the previous section the unit cell anisotropy increases with the dopant concentration and this trend holds for all measured temperatures.

Figure~\ref{fig:3}\textit{c} shows the splitting of the \textit{a}  and \textit{b} lattice parameters which is characteristic of the structural phase transition.  The $x = 0.29$ sample shows the typical behavior of the 122 iron pnictide compounds with the \textit{a} and \textit{b} lattice parameters continuing to diverge with decreasing temperature.  Fitting the order parameter of the structural distortion ($\delta = (a - b)/(a + b)$ shown in Fig~\ref{fig:3}\textit{d}) to a power law of the form $\delta (T) = A_s(T_s-T)^{\beta}/T_s$ the transition temperature can be extracted as well as the critical exponent (see Table~\ref{table:1}). For the $x = 0.29$ sample, a fit critical exponent around  $\beta_s \sim 0.24$ was found, a value very similar to those reported for the \BNFA\ system indicating the similarity between these two systems.\cite{Avci2013}

With increased doping, both the transition temperature and the magnitude of the orthorhombic distortion decrease.  As Sr\textsuperscript{2+} is replaced with Na\textsuperscript{+} the mismatch between the hole and electron pockets increases.  This change in Fermi surface topology weakens the Fermi surface nesting now known to be responsible for the establishment of the antiferromagnetic ordering of the spin-density wave and which, in turn, drives the structural phase transition.\cite{Allred2015N,Castellan2011}  Therefore, the magnitude of the orthorhombic distortion is expected to be related to the strength of the magnetic ordering, Figure~\ref{fig:2}c shows the magnetic moment per Fe site as a function of doping and as expected this parameter decreases, as does the structural distortion, with dopant concentration (magnetism will be more thoroughly discussed in the following section).

For concentrations in the range $0.29  <  x  <  0.42$ the lattice undergos a return to tetragonal symmetry for temperatures below 80K - behavior indicative of the recently discovered magnetic $C_4$ phase.\cite{Avci2014} As can be seen in the temperature dependence of the orthorhombic order parameter, this re-entrant phase is preceded by a suppression of the orthorhombic distortion where the \textit{a} and \textit{b} lattice parameters rapidly converge until $T_r$ at which point the tetragonal \I4mmm\ symmetry is recovered (Figure~\ref{fig:3}\textit{c} and \textit{d}).  This behavior is observed for all samples with compositions in this range  defining a $C_4$ dome with a significantly larger extent in composition space than seen in any previous system.  It is notable that the $x =$ 0.45 sample shows orthorhombic splitting without undergoing tetragonal re-entrance, which describes a $C_4$ dome which closes before the complete suppression of the original $C_2$ SDW phase. Table~\ref{table:2} shows the lattice parameters obtained from Rietveld refinements for a selection of samples at 300 and 10K.

The large thermal contraction along the \textit{c} axis compared to the relatively small shift in the \textit{a} axis at the transition obscures the first-order nature of the transition in variables which measure simultaneously changes in the basal plane and those in the orthogonal direction such as the volume. Introducing the parameter $a_{tet} = \sqrt{a_{oth}^2+b_{orth}^2}/2$) allows for a direct comparision of the tetragonal \textit{a} axis through the transition and clearly shows a lattice anomaly at the transition. In Figure~\ref{fig:4} the temperature dependence of $a_{tet}$ is shown for two representative samples. Here the weakly first-order nature of the first structural transition is clear and an umabiguous anomally in $a_{tet}$ is seen at $T_s$ .  This weakly first-order structural transition is consistent with the observed behavior of both the related hole-doped 122 systems \BKFA\ and \BNFA . \cite{Avci2011,Avci2013}

\subsection{\label{subsect:Tmag}Magnetic properties}

\begin{figure}
\includegraphics[width=\columnwidth]{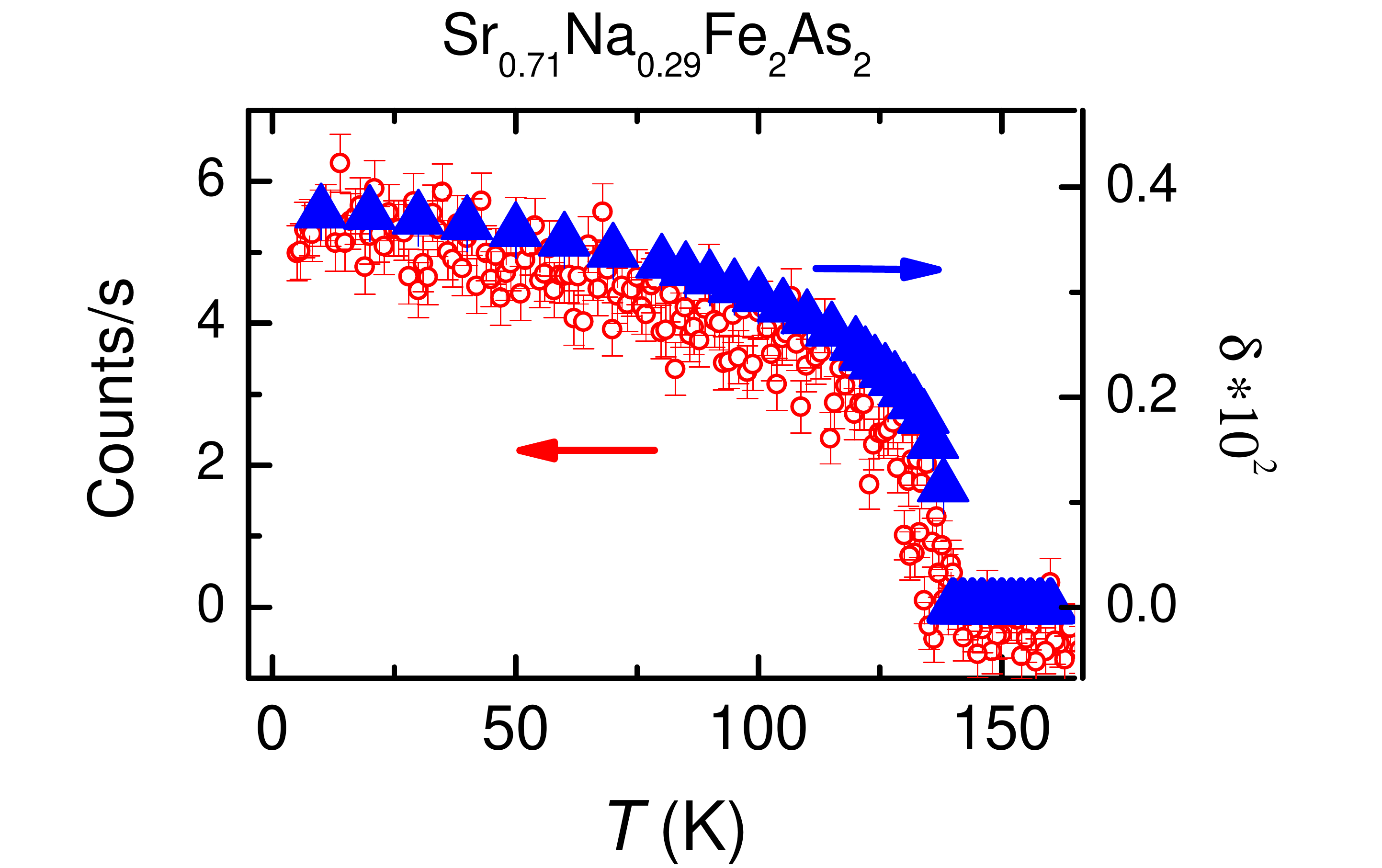}
\caption{\label{fig:5} (Color online) Orthorhomic order parameter and magnetic intensity of magnetic \hht\ peak of $x =$ 0.29 sample scaled and over-plotted to show simultaneity of $T_s$ and $T_N$}
\end{figure}

\begin{figure*}
\includegraphics[width=\textwidth]{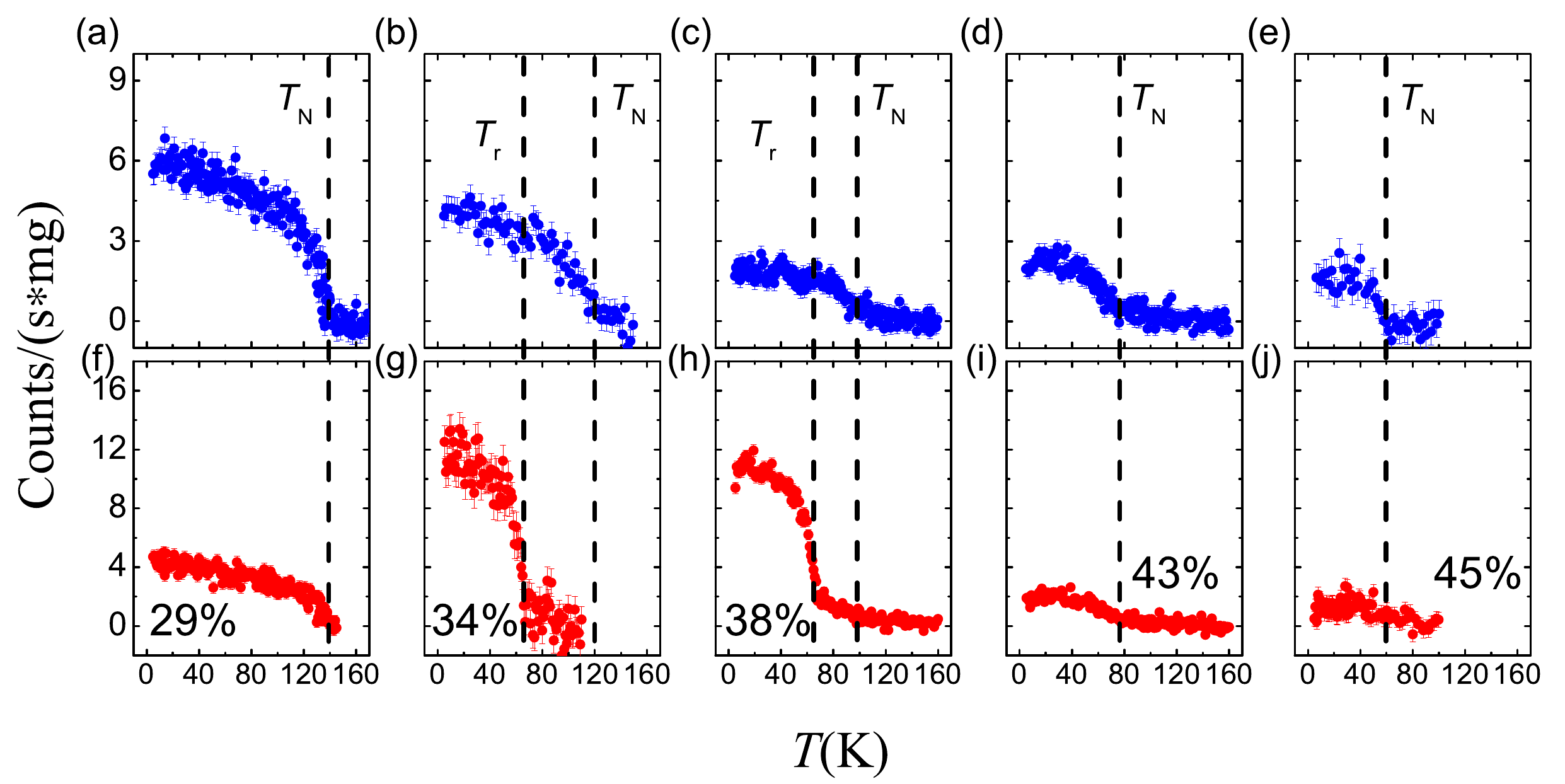}
\caption{\label{fig:6} (Color online) Normalized intensities of magnetic \hho\ (lower panels) and \hht\ (upper panels) reflections for $x =$ 29,34,38,43, and 45\% samples. The typical $C_2$ SDW AFM ordering behavior is seen clearly for the 29\% sample. The magnetic reorientation indicative of the $C_4$ magnetic phase is seen starting with the 34 \% sample and continuing until 43\% where the $C_2$ like behavior is recovered.}
\end{figure*}

In Figure~\ref{fig:5}, the orthorhombic order parameter as determined from structure refinements is over-plotted by the temperature dependence of the \hht\ magnetic peak, both belonging to the $x =$ 0.29 sample. The intensity of the \hht\ magnetic peak follows an effective power-law behavior and can be used as an alternative for the magnitude of the magnetic moment as an order parameter. It is, therefore, useful for the determination of the N\'eel temperature.  Clearly seen is the strong magneto-elastic coupling characteristic to these materials, where the structural distortion and the magnetic moment attain non-zero values simultaneously (on cooling) while demonstrating a similar power-law-like behavior to their temperature dependence.  As described previously in Section~\ref{subsec:Tstruct}, the magnetic intensity can be fit to a power law to obtain $T_N$ as well as the critical exponent, listed in Table~\ref{table:1}.  Comparing the fits for $T_s$ and $T_N$ it is seen that the transitions are simultaneous within the resolution of our experiments.  This is in agreement with the observed first-order character of the transitions in the other members of the hole-doped 122 family where strong magneto-elastic coupling is observed.\cite{Dhital2012,Avci2013,Avci2011,Parshall2015} 

The temperature dependence of the \hho\ and \hht\ magnetic peaks for samples between $0.29 \leq\ x \leq\ 0.45$ are shown in Figure~\ref{fig:6}. Manifest in the \hht\ reflection is the gradual suppression of magnetism upon doping. The \hht\ reflection continually loses intensity as the doping is increased until 48\% (data not shown) where the peak intensity becomes too weak to measure thus defining the edge of the AFM $C_2$ dome.  Simultaneously, the magnetic transition temperature, denoted by the onset of the peak intensity, is seen to decrease, with magnetism ordering at progressively lower temperatures with increased doping.  Both trends are due to the growing mismatch between the hole and electron pockets at the Fermi surface as increasing Na concentrations introduce holes into the electronic structure, resulting in progressively weaker Fermi surface nesting.

\subsection{\label{subsect:C4}Mapping the $C_4$ magnetic phase}    

\begin{figure}
\includegraphics[width=\columnwidth]{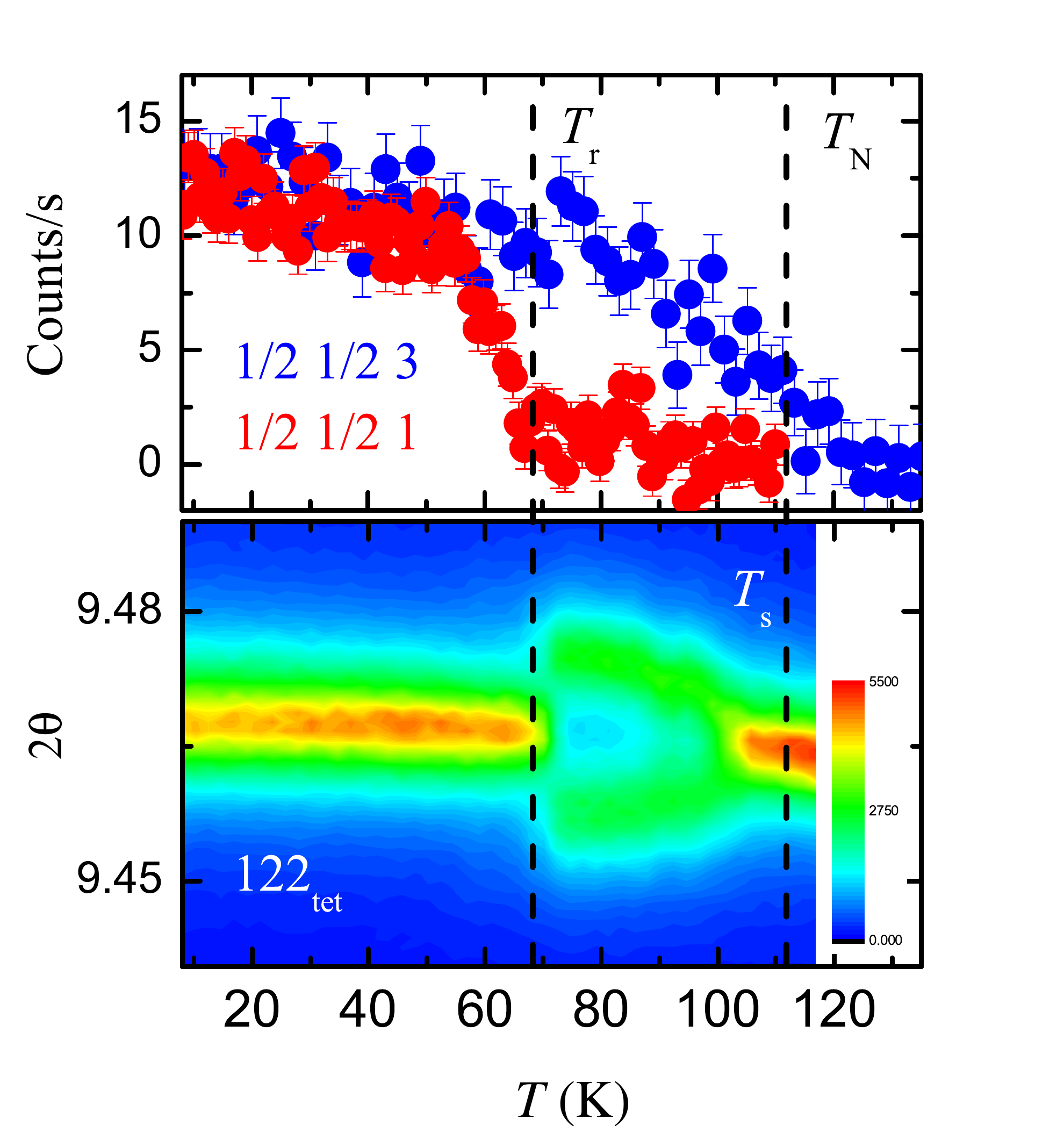}
\caption{\label{fig:7} (Color online) Intensity (upper panel) and diffractogram (lower panel) plots of the \hho\ (red),\hht\ (blue) and 112 peaks respectively.  The 112 peak traces the structural behavior of the lattice splitting at the orthorhombic transition.  The \hho ,\hht\ magnetic peaks show the onset of the SDW as well as the magnetic reorientation assosiated with the $C_4$ phase}
\end{figure}

The observed re-entrance to a tetragonal structural phase for samples with dopings $0.29 < x < 0.45$ is accompanied by a magnetic reorientation which is the hallmark of the magnetic $C_4$ phase.\cite{Avci2014,Allred2015N,Allred2015,Wasser2015} In this phase, the intensity of the \hho\ refection is significantly larger while the intensity of the \hht\ peak is slightly suppressed compared to the well characterized magnetism of the orthorhombic $C_2$ phase.  This behavior can be seen very clearly for the $x =$ 0.34 sample which shows 100\% sample volume re-entrance to the tetragonal phase.  Figure~\ref{fig:7} shows the temperature dependence of both of the magnetic peaks as well as of the nuclear tetragonal 112 peak which splits into the orthorhombic 202 and 022 peaks, of the $x =$ 0.34 sample. From this plot the magnetic reorientation is clear; at the first structural transition magnetic intensity becomes measurable on the \hht\ magnetic peak. Then, at the second structural transition, there is a significant magnetic reorientation and the \hho\ magnetic peak gains more than a factor of three in scattering intensity. This reorientation can be observed in all samples $0.29 < x < 0.43$ as shown in panels a-d and f-i of Fig~\ref{fig:6} which together with the results discussed in Section~\ref{subsec:Tstruct} indicates a robust $C_4$ dome extending over $\Delta x \sim 0.14$ in composition space, considerably larger than that observed in either of the other hole-doped systems with $\Delta x$ being $\sim 0.02$ and $\sim0.04$ for \BKFA\ and \BNFA\ respectively.\cite{Avci2012,Avci2013} 
  
\begin{figure}
\includegraphics[width=\columnwidth]{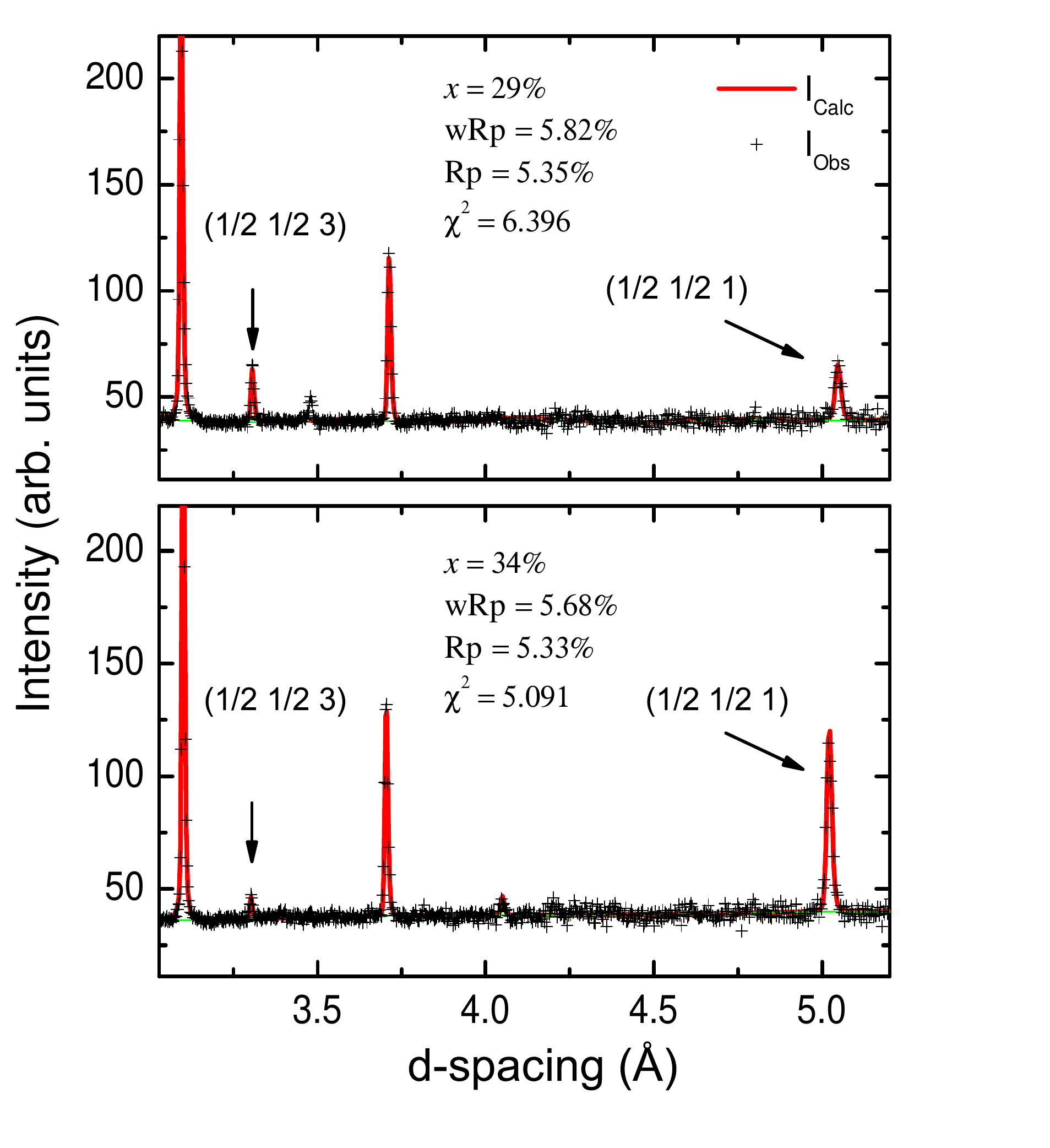}
\caption{\label{fig:8} (Color online) Rietveld fits of orthorhombic and tetragonal magnetic models to $x =$ 0.29 and 0.34 samples. Data collected at 10K on POWGEN. Indicated by arrows are the \hho\ and \hht\ magnetic peaks (as indexed in the \I4mmm\ structure).} 
\end{figure}

Fig~\ref{fig:8} shows representative best-fit Rietveld plots for the $x = 0.29$ and $x = 0.34$ samples obtained from structural refinements performed on patterns collected at 10 K - one sample each from the strictly orthorhombic and re-entrant tetragonal regions described in Section~\ref{subsec:Tstruct}.  While the two samples have different structures at this temperature, both exhibit antiferromagnetic ordering with significant intensities at the \hho\ and \hht\ magnetic peaks and therefore, allow for fitting to different magnetic models.  While the presence of only two or three magnetic peaks makes it practically impossible to converge to a unique magnetic structural model relying solely on neutron powder diffraction, in previous work we proposed two possible models capable of producing satisfactory fits to the tetragonal $C_4$ phase in Ba\textsubscript{$0.76$}Na\textsubscript{$0.24$}Fe\textsubscript{$2$}As\textsubscript{$2$} with both models favoring a spin reorientation from the \textit{ab} plane to the out-of-plane direction, a prediction which has since been confirmed. \cite{Wasser2015,Allred2015}.  The lower panel of Fig~\ref{fig:8} shows the fit of a tetragonal magnetic model with magnetic moments along the \textit{c} axis and $P_C4_2/ncm$ magnetic space group symmetry to the $x =$ 0.34 sample in the $C_4$ region.\cite{Khalyavin2014} This model which forms from the superposition of two magnetic ordering vectors ($\mathbf{Q}_1 = (\pi, 0)$ and $\mathbf{Q}_2 = (0, \pi)$) fits the data well, correctly accounting for the redistribution of magnetic intensity. 

As described in a recent group theoretical analysis work, we determined that the double-\textbf{Q} model necessitates that half the Fe sites become nonmagnetic (nodes) while the remaining half allow the tetragonal SDW antiferromagnetic ordering. \cite{Khalyavin2014,Avci2014}  In later work Wa\ss er,\textit{et al}. used polarized neutron diffraction on a \BNFA\ single crystal exhibiting the $C_4$ phase and found that the magnetic reorientation finds the magnetic moments pointing along the \textit{c} axis. Very recently, we were able to rule in favor of the double Q model using a combination of M\"ossbauer spectroscopy, neutron and synchrotron powder diffraction.\cite{Allred2015N} It is clear that the two AFM phases, hosted on different structures and happening at different temperatures, are both competing with superconductivity.  Interestingly, though the $C_2$ AFM phase is supressed at the onset of superconductivity the magnetic $C_4$ phase in this material clearly supresses superconductivity. \cite{Avci2011,Bohmer2015}  This strong interaction between the double-\textbf{Q} AFM phase and superconductivity is manifest in the nearly flat and low $T_c$ values of $\sim8-10$ K which are only allowed to rise as a function of increasing Na content after leaving the relatively wide $C_4$ dome.  We speculate that the $C_4$ phase might impact the pairing mechanism of the Cooper pairs in the SC phase.

Recent theoretical work and capacitance dilatometry measurements performed on \BNFA\ have suggested the presence of an incommensurate magnetic structure either at the edge of the $C_2$ dome or in the intermediate temperatures $T_r < T < T_N$ for samples exhibiting $C_4$ re-entrance.\cite{Gastiasoro2015,Wang2015} In the analysis presented here no such incommensurate magnetic ordering has been observed. Modeling of the magnetic structure has been performed using high resolution neutron diffraction data collected on POWGEN in each of these regions (see Fig~\ref{fig:8} for example).  Samples with composition $0.29 < x < 0.42$ were well fit by the established $F_cmm^{'}m^{'}$ magnetic space group at temperatures $T > T_r$ with no divergence from the expected peak positions.\cite{Zhao2008,Kaneko2008,Xiao2009,Avci2011,Avci2013} A similar analysis performed on the $x = 0.45$ sample which only shows $C_2$ magnetic structure (see Figure~\ref{fig:6}$e$), also exhibited no observable departure from the known magnetic structure. Therefore, if any incommensurate magnetic ordering in either of these regions is present it must be approximate to the $F_cmm^{'}m^{'}$ to within the resolution of our studies.         

\subsection{\label{subsect:intParam}Internal parameters and the $C_4$ phase}

\begin{figure*}
\includegraphics[width=\textwidth]{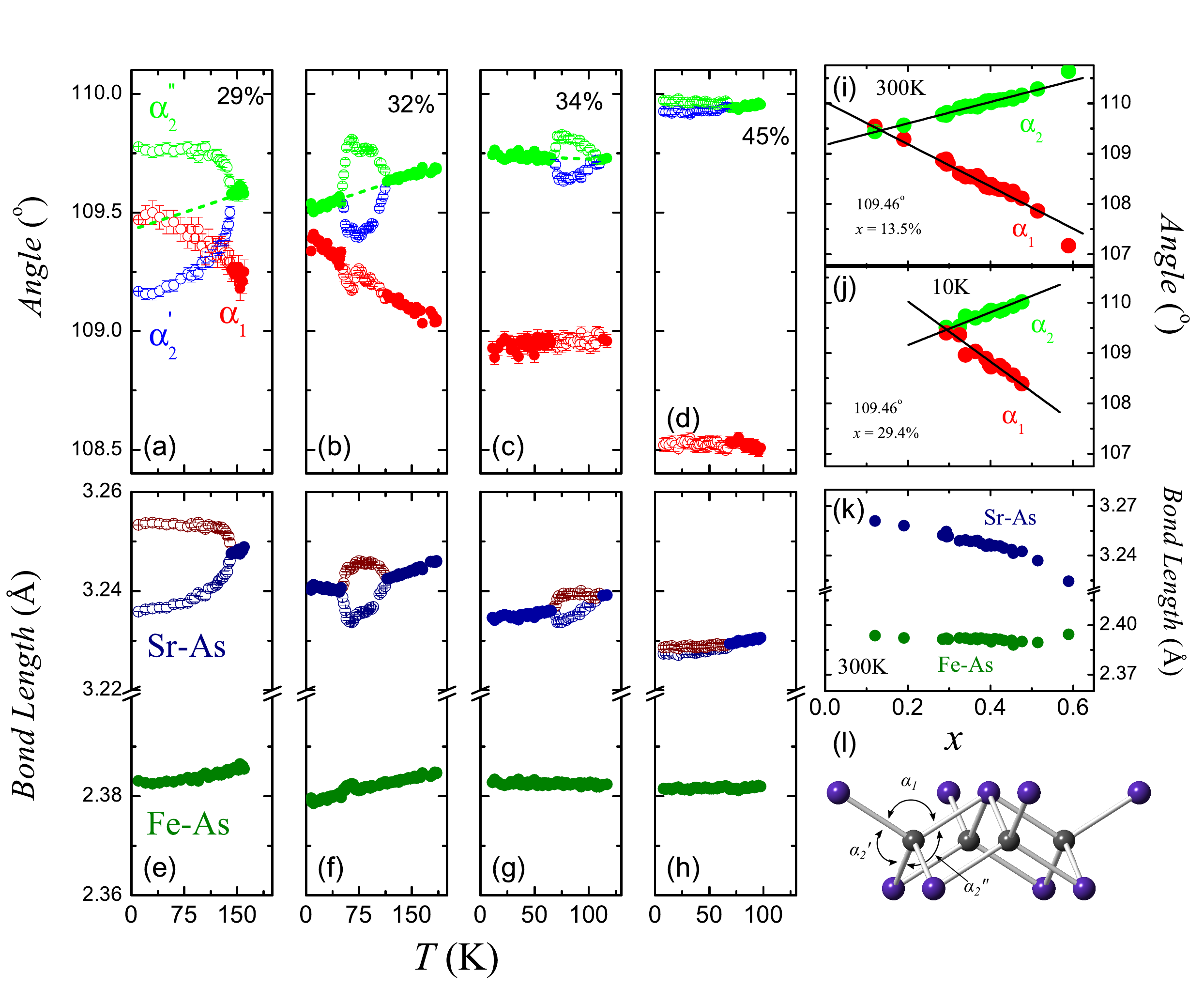}
\caption{\label{fig:9} (Color online) Internal parameters of selected samples. a-d show the temperature dependence of the $\alpha_1$, $\alpha_2$ tetrahedral As-Fe-As angles of the tetragonal phase (filled symbols) while the split $\alpha^{'}_{2}$ and $\alpha^{''}_2$ angles of the orthorhombic phase are denoted with open symbols. Dotted lines have been drawn in for the split angles showing the average behavior. e-h show the Fe-As and Sr-As bond lengths using the same open/filled symbol convention. (i) The room temperature and (j) 10K composition dependence of the angles, where $\alpha^{'}_{2}$ and $\alpha^{''}_2$ have been averaged in (j). (k) The room temperature composition dependence of the Sr-As and Fe-As bond lengths. (l) The Fe\textsubscript{$2$}As\textsubscript{$2$} layer is shown with $\alpha_1$, $\alpha^{'}_{2}$ and $\alpha^{''}_2$ denoted.}
\end{figure*}

In the tetragonal \I4mmm\ symmetry seen in the \AFA\ materials there are six As-Fe-As angles in the Fe\textsubscript{$2$}As\textsubscript{$2$} layers. Due to the symmetry of the structure, these six can be reduced to two related angles defined as $\alpha _1$ and $\alpha _2$. At the structural transition the lowered orthorhombic symmetry causes the $\alpha _2$ angle to split into two separate angles denoted as $\alpha ^{'}_{2} $ and $\alpha ^{''}_{2}$ as shown in Figure~\ref{fig:9}$l$.

Previously (ref~\onlinecite{Avci2013}), we have discussed the competing effects of the Na\textsuperscript{+} ion's smaller ionic radius and reduced electron contribution, when compared to Ba, on bonding in the Fe\textsubscript{$2$}As\textsubscript{$2$} layers and the spirit of these considerations remains unchanged in application to the Sr material. The raised oxidation state of Fe causes a contraction along the Fe-Fe bonds (as directly observed by the contraction of the \textit{a} axis) which, coupled with the rigidity of the Fe-As bond, causes an expansion of the cell along the \textit{c} direction.  

The Fe-As bond rigidity is clearly seen in Fig~\ref{fig:9}\textit{k} where the bond length is essentially constant between $ 0.10 \leq\ x \leq\ 0.50$ at room temperature. This rigidity is also maintained even as a function of temperature as seen in plots e-h where the the Fe-As bond changes by no more than 0.2\% over a temperature range of over 170 K (see Table~\ref{table:2} for bond lengths and angles at 300 and 10K). 

This rather robust rigidity dictates that the previously discussed changes in the lattice and contraction of the Fe-Fe bond must be compensated almost exclusively by the As-Fe-As bond angles and the bonding between the alternating Fe\textsubscript{$2$}As\textsubscript{$2$} and Sr layers. Plotted in Fig~\ref{fig:9}k is the Sr-As bond length which, unlike the Fe-As bond, shows a doping dependence similar to that of the lattice.  While the $\alpha _1$ angle closes with doping (as the \textit{a} axis contracts) the rigidity of the Fe-As bond causes the As atom to be pushed closer to the Sr layer.  A compression of the Sr-As bond compensates for part of this change while still requiring an expansion along the \textit{c} direction as the Fe-As bond becomes more co-linear with the tetragonal axis. Notably, though the Sr-As bond length exhibits significant doping dependence, its average value shows little temperature dependence as seen in panels e-h. However, at the orthorhombic and re-entrant tetragonal transitions the Sr-As bond splits and reunifies similarly to the \textit{a} and \textit{b} lattice parameters, and in the orthorhombic structure there is a significant divergence of the bond lengths saturating at a $\sim 0.02$ \AA\ difference at 10K for the 29\% sample. With little temperature dependence in either the Fe-As or averaged Sr-As bonds the majority of the change in the lattice parameters must be due to the changing of the tetrahedral As-Fe-As bond angles.  

Figure~\ref{fig:9}a-d show the As-Fe-As bond angles as a function of temperature for compositions $x = 0.29, 0.32, 0.34,$ and $ 0.45$. As described above, $\alpha _2$ breaks into two separate angles at the orthorhombic transition as is clearly seen for the $x =$ 0.29 sample.  This allows the $T_r$ and $T_s$ to be tracked in the angle plots and as described in Section~\ref{subsect:C4} the $C_4$ phase is seen for the $x =$ 0.32 and 0.34 samples. Considering the temperature dependence of the angles, it is interesting to note that for the $x =$ 0.29 and 0.32 samples the angles show a strong temperature dependence and quickly either converge or begin to converge as the temperature is lowered, whereas for the $x =$ 0.34 and 0.45 samples the angles are nearly constant over the measured temperature range. Noting the closeness of the angles of the two lower composition samples to $109.46^\circ$, it is tempting to ascribe this behavior to a special preference of this structure to this angle. Considering other analyses presented above, it seems unlikely and unsupported that the lattice is significantly more sensitive to the introduction of Na near the parent compound than at higher dopings - the lattice anisotropy is linear in composition throughout this range - and therefore effects other than just the contraction of the lattice due to substitution must be contributing; however, more work is needed to fully understand this behavior. 

Panels i and j show the the two As-Fe-As angles ($\alpha _1$ and $\alpha _2$) at 300 and 10K respectively (at 10K $\alpha _2$ is the average of $\alpha _{2} ^{'}$ and $\alpha _{2} ^{''}$ for the compositions with the orthorhombic structure). Unlike the other hole-doped Ba systems the smaller Sr series with the lower cell anisotropy (as measured by \textit{c/a} and discussed in Section~\ref{subsect:dopdep}) starts with an end member already close to the perfect tetrahedral angle of $109.46^\circ$.  Upon doping the larger $\alpha _1$ closes while $\alpha _2$ opens until $x \sim 13\%$ where $109.46^\circ$ is achieved and $\alpha _2$ becomes the larger angle with further doping. Comparing plots i and j it is clear that the composition at which this angle is achieved is significantly affected by the temperature: it changes from $x_{109.46^\circ} = 0.135$ to $x_{109.46^\circ} = 0.294$ between 300 and 10 K. It is interesting that despite the non-linear behavior of the lattice volume the doping dependence of the averaged angles appears linear across all measured compositions. 

The perfect tetrahedral angle at 10 K occurs at a composition near that of the start of the magnetic $C_4$ phase. It is likely that the proximity to the higher symmetry perfect tetrahedron might give way to a structural instability which leads to the re-establishment of the tetragonal structure. Calculating the $\Delta \alpha = \| \alpha_2-\alpha_1 \|$ for all samples which show $C_4$ re-entrance in both the Na doped \BFA\ (taken from reference \onlinecite{Avci2013}) and SrFe\textsubscript{$2$}As\textsubscript{$2$} it is found that only compositions with $\Delta \alpha \lesssim 1^\circ$ show the magnetic reorientation and the structural re-entrance. We suggest that a combination of the magnetic ordering temperature of a given composition and its proximity to $\Delta \alpha = 0$ should play a role (along with considerations of the Fermi surface) in establishing the $C_4$ magnetic phase - explaining why the $C_4$ phase is not seen to extend to $\Delta\alpha \geq 1^{\circ}$ for compositions $x < 0.294$.  The ability of the high $C_2$ SDW ordering temperatures to supress the formation of the $C_4$ phase has been corroborated by recent theoretical work, supporting this conjecture. \cite{Kang2015}

\begin{figure}
\includegraphics[width=\columnwidth]{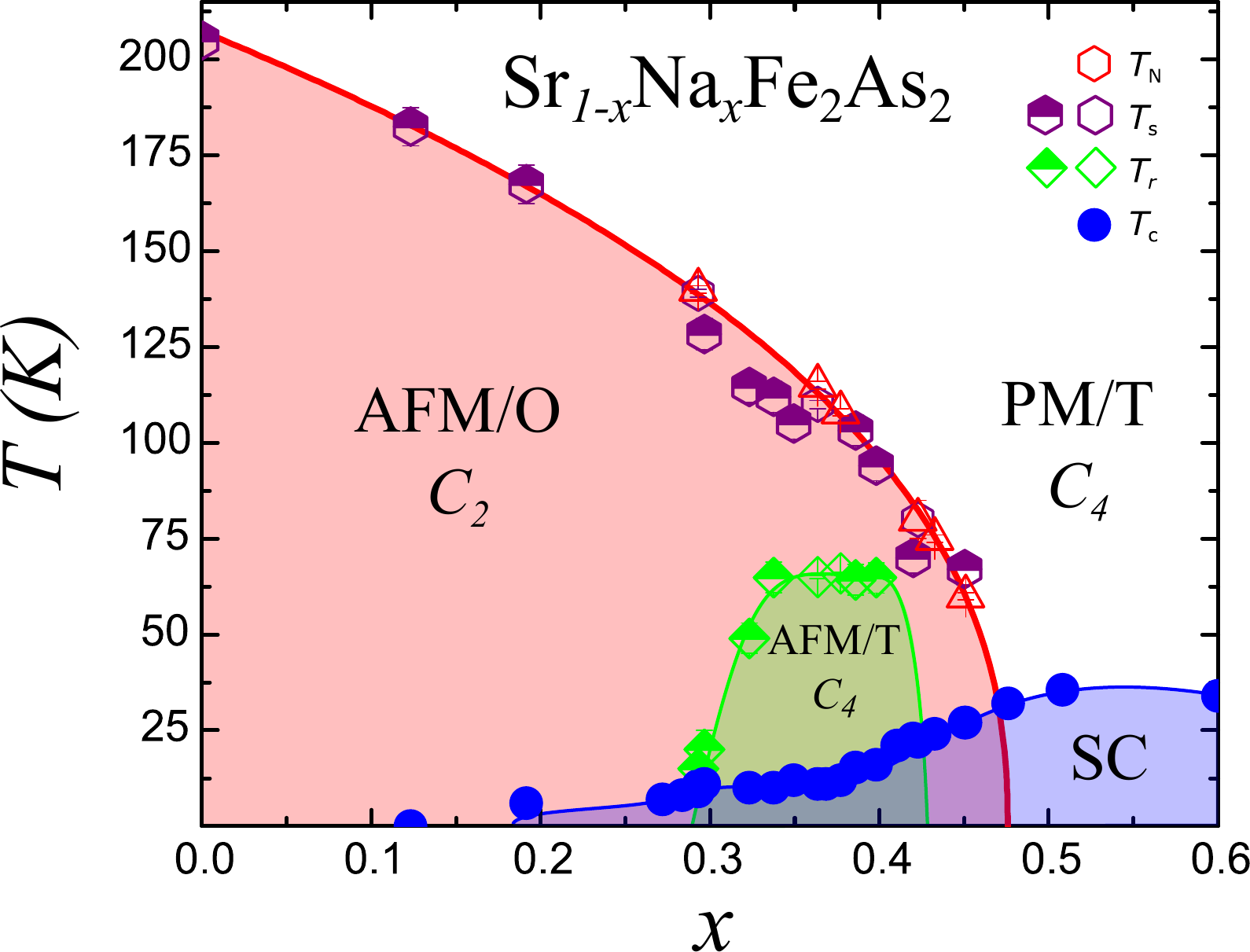}
\caption{\label{fig:10} (Color online) Phase diagram of \SNFA\ transition temperatures determined from x-ray and neutron diffraction are denoted by half filled and empty shapes respectively. PM/T $C_4$ is the normal state paramagnetic tetragonal phase while AFM/T $C_4$ is the magnetic tetragonal phase. AFM/O $C_2$ is the orthorhombic antiferromagnetic phase. \lq SC\rq\ denotes superconducting samples independent of phase structure. }
\end{figure}

\section{\label{sec:conc}Conclusions}

Figure~\ref{fig:10} shows the culmination of all discussions presented above in a \SNFA\ phase diagram. In contrast to previously published phase diagrams we have observed a robust $C_4$ dome which spans $\Delta x = 14\%$ in composition space, stabilizes at the high temperature of 65 K, and whose extent is entirely within the $C_2$ dome - closing before the complete suppression of the SDW AFM ordering.\cite{Cortes2011,Shinohara2015} 

Interestingly, the re-entrance of the $C_2$ SDW phase as the ground state at compositions near $T_N \rightarrow 0$ reported here was predicted in recent theoretical work ref~\onlinecite{Kang2015} which endeavored to recreate the features observed in the phase diagram produced in ref~\onlinecite{Bohmer2015}. In their mean-field approximation Jang \text{et al}. show that, for large dopant concentrations which also display low $T_N$, the energies of the $C_4$ and $C_2$ SDW states become nearly equivalent, with the $C_2$ structure being slightly more energetically favorable. While not a feature observed by B\"ohmer \textit{et al}. in \BKFA\ or in our previous work on \BNFA\ (see ref~\onlinecite{Avci2014}) here, for \SNFA , the behavior is seen as a clear separation of $\Delta x \sim 0.03$ between the closing of the $C_4$ and $C_2$ domes (Fig~\ref{fig:10}). We attribute this difference between the hole-doped systems to the higher AFM ordering temperature seen in the SrFe\textsubscript{$2$}As\textsubscript{$2$} parent material, which allows the SDW dome to persist to higher dopant concentrations and thus to compositions for which this near degeneracy between the two SDW phases occurs.        

As recently reported for the \BKFA\ system (refs~\onlinecite{Bohmer2015,Allred2015}), we observe a strong competition between the $C_4$ state and superconductivity.  As opposed to the smoothly sloping superconducting dome archetypical to the iron pnictides we report a large plateau in the $T_c$ at the onset of the $C_4$ magnetic phase where the superconducting transition is nearly constant until the end of the $C_4$ dome where it immediately begins to climb to its maximum value of 36K for $x=0.49$. Whereas reference \onlinecite{Bohmer2015} observed non-monotonic doping dependence to $T_c$ in \BKFA\ we see no such decrease at the onset of $C_4$. We report coupled simultaneous magnetic and structural transitions both at the well known structural and magnetic transitions to an orthorhombic AFM structure and at the newly observed magnetic reorientation and tetragonal re-entrance, in general agreement with the strong magneto-elastic coupling in the hole-doped iron pnictide 122 materials.     

While in both the K and Na doped \BFA\ the $\alpha = 109.46^\circ$ occurs  near the composition of optimum $T_c$ here it is achieved well inside the SDW dome. It is possible that this observation was simply a coincidence in the previous two systems, which we think unlikely, or that in light of the well known competition between the two magnetic phases and superconductivity the maximum transition temperature is not found at the perfect tetrahedral angle due to the strong magnetic ordering present in this composition.  

With the observation of the magnetic $C_4$ phase in a third member of the hole-doped 122 iron pnictide superconductors it strongly suggests its universality to these systems. Moreover, it suggests that this new phase is important to the wider material group and not just an isolated observation.

\begin{acknowledgments}
The work at the Materials Science Division at Argonne National Laboratory was supported by the US Department of Energy, Office of Science, Materials Sciences and Engineering Division. The part of the research that was conducted at the ANL Advanced Photon Source and at the ORNL High Flux Isotope Reactor and Spallation Neutron Source was sponsored by the Scientific User Facilities Division, Office of Basic Energy Sciences, US Department of Energy. The authors thank A. Huq, P. Whitfield and A.A. Aczel  for providing help during experimental collection and analysis.
\end{acknowledgments}

\end{document}